\documentclass[
  journal=nws,
  manuscript=research-article,
  year=2025,
  volume=0,
]{cup-journal}

\usepackage{amsmath,amsfonts,amssymb}
\usepackage{graphicx}
\usepackage{subcaption}
\usepackage{hyperref}
\usepackage{xcolor}
\usepackage{float}
\usepackage{array}
\usepackage{booktabs}
\usepackage{quotes}
\usepackage{listings}
\usepackage{xcolor}

\definecolor{goodblue}{RGB}{0, 91, 187}
\hypersetup{
  colorlinks=true,
  allcolors=goodblue,
  urlcolor=goodblue,
  citecolor=goodblue,
  pdfborder={0 0 0},
  breaklinks=true,
}

\lstdefinelanguage{json}{
  basicstyle=\ttfamily\tiny,
  stepnumber=1,
  numbersep=10pt,
  showstringspaces=false,
  breaklines=true,
  frame=single,
}

\lstdefinestyle{python}{
    language=Python,
    basicstyle=\ttfamily\small,
    numberstyle=\tiny\color{gray},
    stepnumber=1,
    numbersep=10pt,
    backgroundcolor=\color{gray!10},
    showstringspaces=false,
    tabsize=4,
    captionpos=b,
    breaklines=true,
    keywordstyle=\color{blue}\bfseries,
    commentstyle=\color{green!50!black},
    stringstyle=\color{magenta},
    frame=single,
    rulecolor=\color{black!30},
    frameround=tttt
}

\lstdefinestyle{R}{
    language=R,
    basicstyle=\ttfamily\small,
    numberstyle=\tiny\color{gray},
    stepnumber=1,
    numbersep=10pt,
    backgroundcolor=\color{gray!10},
    showstringspaces=false,
    tabsize=2,
    captionpos=b,
    breaklines=true,
    keywordstyle=\color{blue}\bfseries,
    commentstyle=\color{green!50!black},
    stringstyle=\color{magenta},
    frame=single,
    rulecolor=\color{black!30},
    frameround=tttt
}

\lstdefinelanguage{Julia}{
  morekeywords={
    abstract, break, catch, const, continue, do, else, elseif, end, export,
    false, for, function, global, if, import, in, let, local, macro, module,
    mutable, primitive, quote, return, struct, true, try, type, using, while
  },
  sensitive=true,
  morecomment=[l]\#,
  morestring=[b]",
  morestring=[b]'
}
\lstdefinestyle{julia}{
    language=Julia,
    basicstyle=\ttfamily\small,
    numberstyle=\tiny\color{gray},
    stepnumber=1,
    numbersep=10pt,
    backgroundcolor=\color{gray!10},
    showstringspaces=false,
    tabsize=4,
    captionpos=b,
    breaklines=true,
    keywordstyle=\color{blue}\bfseries,
    commentstyle=\color{green!50!black},
    stringstyle=\color{magenta},
    frame=single,
    rulecolor=\color{black!30},
    frameround=tttt
}

\newcommand{\us}{\_\;\!}

\title{HIF: The hypergraph interchange format for higher-order networks}

\author{Mart\'in Coll}
\affiliation{Department of Computer Science, University of Buenos Aires, CABA, C1428EGA, Argentina}

\author{Cliff A.\ Joslyn}
\affiliation{Pacific Northwest National Laboratory, Seattle, 98109, Washington, USA}
\alsoaffiliation{Department of Systems Science and Industrial Engineering, Binghamton University, Binghamton, 13902, New York, USA}

\author{Nicholas W. Landry}
\affiliation{Department of Biology, University of Virginia, Charlottesville, 22904, Virginia, USA}
\alsoaffiliation{School of Data Science, University of Virginia, Charlottesville, 22904, Virginia, USA}
\alsoaffiliation{Vermont Complex Systems Institute, University of Vermont, Burlington, 05405, Vermont, USA}
\email[Nicholas W. Landry]{nicholas.landry@virginia.edu}

\author{Quintino Francesco Lotito}
\affiliation{Department of Information Engineering and Computer Science, University of Trento, Trento, 38123, Italy}
\alsoaffiliation{Department of Network and Data Science, Central European University, 1100 Vienna, Austria}

\author{Audun Myers}
\affiliation{Pacific Northwest National Laboratory, Seattle, 98109, Washington, USA}

\author{Joshua Pickard}
\affiliation{Eric and Wendy Schmidt Center, Broad Institute, Cambridge, 02142, Massachusetts, USA}
\alsoaffiliation{Department of Computational Medicine \& Bioinformatics, University of Michigan, Ann Arbor, 48109, Michigan, USA}

\author{Brenda Praggastis}
\affiliation{Pacific Northwest National Laboratory, Seattle, 98109, Washington, USA}

\author{Przemysław Szufel}
\affiliation{SGH Warsaw School of Economics, Warsaw, 02-554, Poland}

\keywords{network, higher-order, hypergraph, directed hypergraph, simplicial complex, data standard} 

\begin{document}

\begin{abstract}
Many empirical systems contain complex interactions of arbitrary size, representing, for example, chemical reactions, social groups, co-authorship relationships, and ecological dependencies.
These interactions are known as higher-order interactions and the collection of these interactions comprise a higher-order network, or hypergraph.
Hypergraphs have established themselves as a popular and versatile mathematical representation of such systems and a number of software packages written in various programming languages have been designed to analyze these networks.
However, the ecosystem of higher-order network analysis software is fragmented due to specialization of each software's programming interface and compatible data representations.
To enable seamless data exchange between higher-order network analysis software packages, we introduce the Hypergraph Interchange Format (HIF), a standardized format for storing higher-order network data.
HIF supports multiple types of higher-order networks, including undirected hypergraphs, directed hypergraphs, and abstract simplicial complexes, while actively exploring extensions to represent multiplex hypergraphs, temporal hypergraphs, and ordered hypergraphs.
To accommodate the wide variety of metadata used in different contexts, HIF also includes support for attributes associated with nodes, edges, and incidences.
This initiative is a collaborative effort involving authors, maintainers, and contributors from prominent hypergraph software packages.
This project introduces a JSON schema with corresponding documentation and unit tests, example HIF-compliant datasets, and tutorials demonstrating the use of HIF with several popular higher-order network analysis software packages.
\end{abstract}

\section{Introduction}

The growth of network science as a field has been fueled by the availability of large-scale datasets from a variety of sources including the internet~\autocite{albert_diameter_1999}, co-authorship networks~\autocite{barabasi_evolution_2002,newman_structure_2001}, global-scale transportation~\autocite{brockmann_hidden_2013}, social interactions~\autocite{cattuto_dynamics_2010,stehle_highresolution_2011,adams_gathering_2019}, biological processes~\autocite{barabasi_network_2004,jeong_largescale_2000}, social media~\autocite{jackson_hashtagactivism_2020,morstatter_sample_2013,ugander_structural_2012}, and brain connectivity~\autocite{sporns_organization_2004}, among others.
As the plethora of data has grown, standardized formats for pairwise network datasets have been developed such as GraphML~\autocite{brandes_graph_2010} and Pajek~\autocite{nooy_exploratory_2018} as well as network dataset repositories such as SNAP~\autocite{leskovec_snap_2014}, Netzschleuder~\autocite{peixoto_netzschleuder_2021}, ICON~\autocite{clauset_colorado_2016}, and Konect~\autocite{kunegis_konect_2013}.
These advances, however, have been developed with networks primarily in mind, which solely model dyadic interactions.

Higher-order networks generalize the notion of networks, representing interactions of arbitrary size.
These networks can more naturally model certain empirical systems such as co-authorship networks, chemical reactions, human social networks, and biological interactions, which abound in multi-way interactions.
This can be useful for studying nonlinear dynamical processes~\autocite{neuhauser_multibody_2020}, the evolution of groups~\autocite{iacopini_temporal_2024}, and protein-protein interactions~\autocite{murgas_hypergraph_2022}, for example.
There are many robust software packages for storing, analyzing, and visualizing higher-order networks~\autocite{failla_attributed_2023,pickard_hat_2023,diaz_hypergraphsjl_2022,lotito_hypergraphx_2023,praggastis_hypernetx_2024,badie-modiri_reticula_2023,spagnuolo_analyzing_2020,hajij_topox_2024,landry_xgi_2023}, but no corresponding data standards for facilitating cross-package compatibility.
In this paper, co-authored by leaders and representatives of a number of these software packages, we explain the need for such a standard and the context in which it fits; introduce the Hypergraph Interchange Format (HIF) as a data standard for higher-order networks;  describe integration with existing higher-order network software libraries; and lastly, demonstrate how it can be used to unlock the strengths of different higher-order network software packages.

\subsection{Existing higher-order network data}

Higher-order network data remains relatively scattered, with multiple versions of the same datasets often distributed across several repositories and lacking a consistent format.
For example, the webpage curated by Austin Benson~\autocite{benson_data_2021} features many datasets, primarily from the computer science domain, but is no longer actively maintained.
There exist several actively maintained data repositories: XGI-DATA~\autocite{landry_xgi_2023}, which hosts many datasets on Zenodo \footnote{See \href{https://zenodo.org/communities/xgi}{https://zenodo.org/communities/xgi}.} and is integrated with the XGI software API; HypergraphRepository~\autocite{antelmi_hypergraphrepository_2024}, which offers a collection of empirical hypergraph datasets along with a web application for filtering and downloading them; and the Hypergraphx~\autocite{lotito_hypergraphx_2023} software library, which is accompanied by a diverse collection of datasets spanning multiple domains and hypergraph types~\autocite{HGXTeamData}.
Additionally, higher-order network datasets can sometimes be found in traditional network repositories such as SNAP~\autocite{leskovec_snap_2014}, ICON~\autocite{clauset_colorado_2016}, and Netzschleuder~\autocite{peixoto_netzschleuder_2021}, typically represented as bipartite graphs (see discussion in Section \ref{sec:hon} below). 
Notably, there may be repetitions and overlapping datasets across these repositories, resulting in a redundant and fragile ecosystem.
Furthermore, each repository currently uses its own proprietary format, which is not directly interoperable with other software libraries.

As research and development in hypergraph-based methods expand across diverse domains such as data mining, computational biology, and network analysis, the absence of a common data interchange format has become an obstacle to interoperability and reproducibility.
Most existing libraries and tools rely on ad hoc representations of hypergraphs, making it difficult to share data, compare results, or construct integrated workflows.
In practical applications, hypergraphs typically pass through multiple stages of analysis such as construction, transformation, and visualization, with each stage potentially handled by a different tool.
Without a standardized format, these transitions require custom conversion scripts and increase the risk of semantic loss or data misinterpretation.

\subsection{Contributions}

HIF addresses this challenge by offering a consistent, expressive, and extensible JSON-based schema for encoding both the topology and metadata of hypergraphs.
Its language-agnostic design enables seamless exchange between libraries implemented in Python, C++, Julia, JavaScript, and other environments, promoting modular, interoperable workflows while reducing redundant engineering effort.
HIF serves as a unifying layer for the higher-order network science ecosystem, allowing software tools and pipelines to communicate through a shared representation.
This facilitates reproducible research and fosters a more cohesive, collaborative development environment, similar to the role of the Open Neural Network Exchange (ONNX) format in machine learning~\autocite{onnx2024} and GraphML in traditional graph processing~\autocite{brandes_graph_2010}.

\section{\label{sec:hon} Higher-Order Networks as Hypergraphs}

Network science is grounded on the mathematical representation of networks as undirected or directed graphs. Similarly the term ``higher-order network'' has become associated with a variety of data objects related to undirected and directed hypergraphs.
The literature on hypergraphs is extensive, and we refer the reader to two classic publications about hypergraphs~\autocite{berge_hypergraphs_1984,bretto_hypergraph_2013} as well as more recent treatments of the larger space of higher-order networks~\autocite{battiston_networks_2020, torres_why_2021a,joslyn_hypernetwork_2021, bick_what_2023a}.

The foundational mathematics of hypergraphs continues to evolve, including in a category-theoretical framework~\autocite{grilliette_incidence_2023}, producing multiple ways to axiomatize hypergraph formalisms. Our approach for this paper is to seek an appropriate general grounding sufficient to accommodate the widest range of both current hypergraph modeling approaches and current data science needs, while avoiding some of the more subtle complexities which can arise especially around edge cases like empty and duplicate hyperedges, isolated and redundant vertices, empty hypergraphs, self-loops, and related things. 
To that end, we now introduce just a few concepts of the most direct relevance to HIF formally, while noting both similarities and differences with some of the most widely circulated concepts as used in the literature.

Thus for our purposes, a {\bf hypergraph} is a system $H=(V, E, I)$ where $V= \{v\}$ is a finite, non-empty set of {\bf vertices} or {\bf nodes}, $E=\{e\}$ is a finite, non-empty set of {\bf edges} or {\bf hyperedges},\footnote{We will commonly refer to ``edges'' to mean hyperedges for convenience and when clear from context.} and $I \subseteq V \times E$ is a set of {\bf incidences}, that is, pairs $(v, e)$ of nodes and edges. 

It may be valuable to understand how this definition may differ from some readers' expectations.
First, note that an edge $e \in E$ can be mapped to the collection of vertices with which it has an incidence: $e \mapsto \{ v \in V : (v, e) \in I \}$.
With this in mind, we will also feel free to simply think of an edge as a subset of vertices, and notate $e \subseteq V$, thereby recovering the traditional sense of a hyperedge as a collection of vertices of arbitrary size.
And indeed, we could also notate $H=(V, E)$ and simply understand that each edge $e \in E$ is a subset $e \subseteq V$.
But what's critical to note is that then this $E$ is not actually a set, but rather, using our definition $E$ would be a multiset or  bag of hyperedges, and so possibly containing duplicates: there can be two distinct edges $e, f \in E$ which map to the {\em same} set of vertices, so that $e=f$.
So while mathematically, our structure may be properly called a {\bf multi-hypergraph}, for this paper, we will simply call these all hypergraphs, while still keeping this issue firmly in mind.

Rather more well understood is how any particular edge is a subset $e \subseteq V$ of arbitrary size.
If $e$ maps to only two vertices, then that $e$ is a traditional (undirected) graph edge; and indeed, if this is true for {\em all} edges $e \in E$, then $H$ is called a {\bf 2-uniform hypergraph}, which is just a {\bf graph}: all graphs are hypergraphs.
But $e$ may also have more than two nodes (the traditional sense of "higher-order"), one node (singleton hyperedges), or even none (empty hyperedges are permitted).

HIF supports {\bf directed hypergraphs}~\autocite{ausiello_directed_2001,gallo_directed_1993a}, which comprise directed hyperedges $e \in E$. Each directed hyperedge is an ordered pair $e=(t,h)$ of subsets of vertices: the ``tail'' $t \subseteq e$ specifies the inputs in the multi-way interaction, and the "head" $h \subseteq e$, specifies the outputs.
While the tail and head cover the hyperedge in that $t \cup h = e$, in contrast to some researchers ~\autocite{jost_hypergraph_2019} who require that the head and tail be disjoint, here we generalize this framework and allow nodes to belong to both the tail and head.

Another prominent structure in higher-order network theory is that of an {\bf abstract simplicial complex (ASC)}.
An ASC is a hypergraph, but one which is "downward closed" in that for every hyperedge $e \in E$, then all nonempty collections of vertices $f \subseteq e$ contained in $e$ are also in the hypergraph: $f \in E$. ASCs are used extensively in higher-order network science (\cite{joslyn_hypernetwork_2021}), for example when we seek to analyze the topological properties of hypergraphs mathematically as structures with multiple multi-dimensional components glued together, or when semantically it is understood in some application that when a relationship holds for a set of entities, it also holds for all its subsets.

Essential to HIF is that it supports {\bf attributed hypergraphs}: nodes $v \in V$ and edges $e \in E$ can have arbitrary properties.
But of great significance is that in addition, HIF also supports properties on the {\em incidences} $(v, e) \in I$.
In other words, a node $v \in V$ can have one property when associated with an incidence $(v, e)$ involving one edge $e \in E$ in which it is contained ($v \in e$), but could have quite a different property when associated with a different incidence $(v, f)$ for a different edge $f \in E$, for which also $v \in f$. 
An incidence property can be used to represent directed hyperedges, in that for any particular hyperedge $e \in E$ and vertex $v \in e$, an incidence property can be used to encode whether $v$ is in the tail $v \in t$, the head $v \in h$, or in both.
Incidence properties are also essential for supporting critical features like "edge-dependent weights"~\autocite{chitra_random_2019} used in, e.g., random walk models on hypergraphs~\autocite{hayashi_hypergraph_2020}. 

It is also worth noting a few additional properties of hypergraphs and mathematical structures related to hypergraphs. While these aren't represented in HIF proper, they are important for applications and understanding, and are mentioned elsewhere in this paper in the context of their implementation in some of the hypergraph software systems supporting HIF.

First, the incidence relation $I \subseteq E \times V$ is commonly represented as a Boolean {\bf incidence matrix}, sometimes also notated (perhaps abusively) as just $I$, with $|V|$ rows and $|E|$ columns, where $I[i,j] = 1$ when $(v,e) \in I$ and $0$ when $(v,e) \not \in I$. Matrix operations on $I$ can be very useful in applications, as we will see below. Additionally, every hypergraph $H = (V,E,I)$ has an equivalent {\bf dual} form $H^* = (E,V,I^{-1})$ where nodes and edges are swapped, and $I^{-1}$ is the inverse of the incidence relation, yielding the dual incidence matrix $I^T$ as the transpose. 

Every hypergraph $H$ is also equivalently represented as a {\bf bipartite graph} $B=( V \sqcup E, A )$ on a new set of vertices as the disjoint union of the nodes $V$ and hyperedges $E$ of the original hypergraph, and with a new set of graph edges $A \subseteq \binom{V \sqcup E}{2}$, but with the bipartite condition that for all edges $\{x,y\} \in A$, exactly one of $x$ or $y$ are in $V$, with the other in $E$. Undirected hypergraphs yield undirected bipartite graphs; directed hypergraphs yield directed bipartite graphs; and vice versa in both cases. While hypergraphs are bijective with bipartite graphs, they can represent different concepts and functions, and can be amenable to different algorithms: hypergraphs are about connectivity and path following, where bipartite graphs are about combinatoric and matching questions. And from a mathematical perspective, they may also have different transformations which yield them in different categories (\cite{grilliette_incidence_2023}).

Hypergraphs operate in a broader mathematical ecosystem which also includes graphs. We have already seen how every graph is a 2-uniform hypergraph. Additionally, given a hypergraph $H=(V,E,I)$, it is common to construct a graph $(V,C)$ on its vertex set, called a {\bf clique expansion}, {\bf 2-section}, or {\bf underlying graph}. Here  $C$ is the union of the complete graphs (cliques) formed on each hyperedge $e \in E$. Thereby $C$ consists of all the pairs of vertices (graph edges on $V$) included in some hyperedge, such that for all $\{u,v\} \in C, \exists e \in E, \{u,v\} \subseteq e$. Note that the adjacency matrix of the 2-section is given by $I \times I^T$ as a matrix operation. Finally, it is also common to build the {\bf line graph} of a hypergraph $H=(V,E,I)$ as a graph $(E,L)$ on the edges taken as vertices, where a pair of edges $\{e,f\}$ is included in $L$ if they intersect: $L = \{ e,f \in E : e \cap f \neq \emptyset \}$. The line graph structure allows reasoning about the relationships between the hyperedges, as opposed to the vertices, and is thus closely related to the dual hypergraph $H^*$: indeed, the line graph of $H$ is just the 2-section of the dual $H^*$, and vice versa, where now the adjacency matrix of the line graph is $I^T \times I$.

\section{The Hypergraph Interchange Format}

The Hypergraph Interchange Format is a standard for storing higher-order networks in JavaScript Object Notation (JSON) format.
JSON is chosen for its interpretability; while there are more efficient formats such as Parquet, CSV, GraphQL, etc., these data structures either lack human-readability or support for the rich attributes often accompanying higher-order data.
JSON is supported by all major programming languages, and JSON validation libraries written in Python, R, and Julia provide a user-friendly way to verify datasets, as seen in Figure~\ref{fig:schema_validation_code}, making this a suitably popular file format.
HIF offers support not only for network-level metadata, but also for attributes corresponding to nodes, hyperedges, and incidences, and it does so in an interpretable way.
This specification also handles several different types of higher-order networks and is flexible enough to accommodate additional higher-order representations in the future.

HIF is the first cross-platform standardized file format for datasets of higher-order networks, hoping to unify the higher-order network science community around a set of standards and best practices.
This standard was crystallized through exploration of the practical needs of the community and limitations of implementation.
Not only do we describe a data standard, but we also provide a workflow for updating this schema as the needs of the higher-order network science community evolve and JSON schema specifications are updated.

\subsection{The Hypergraph Interchange Format specification}

\begin{figure}
\begin{subfigure}[b]{0.45\textwidth}
    \includegraphics[width=\textwidth]{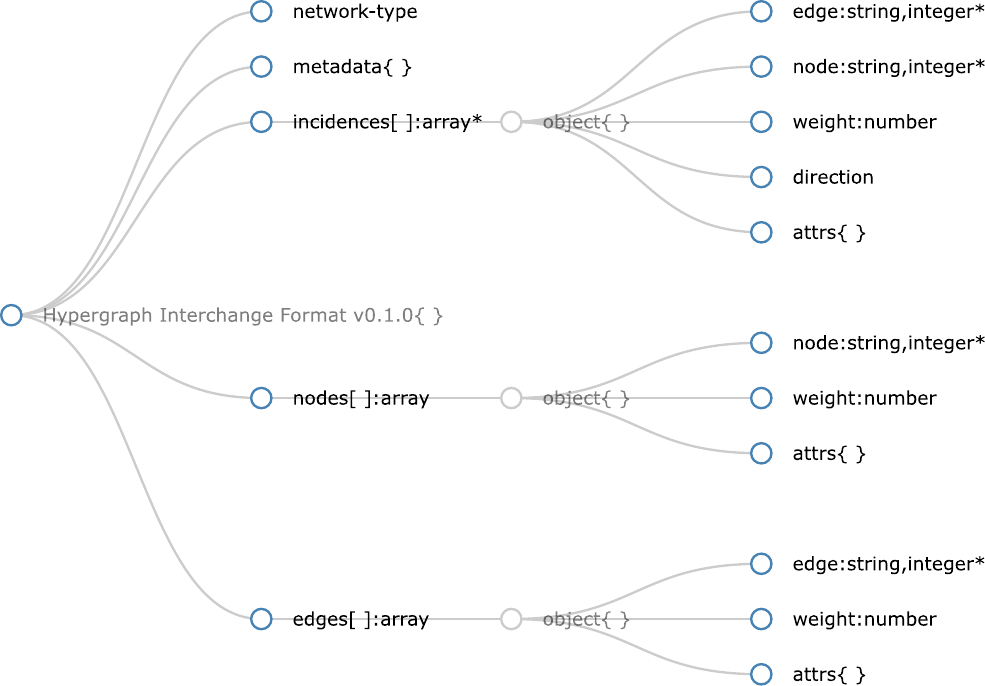}
    \caption{The hierarchy of properties in the HIF JSON Schema.}
\end{subfigure}
\hspace{1cm}
\begin{subfigure}[b]{0.45\textwidth}
\begin{lstlisting}[language=json]
{
    "network-type": "undirected",
    "metadata": {
        "title": "HIF publications dataset",
        "collection-date": "2025-07-04",
        "doi": "10.5281/zenodo.15802760"
    },
    "incidences": [
        {"edge": 10, "node": 20, "weight": 2},
        {"edge": 30, "node": 20, "weight": 3}
    ],
    "nodes": [
        {"node": 20, "attrs":
            {"institutions": ["PNNL", "UVA"]}}
    ],
    "edges": [
        {"edge": 10, "attrs":
            {"source": "Arxiv",
             "tags": ["Machine Learning", "Artificial Intelligence"]}},
        {"edge": 30, "attrs":
            {"source": "DBLP", "tags": ["Hypergraph"]}}
    ]
}
\end{lstlisting}
\vspace{-0.25cm}
\caption{An example HIF document in JSON format.}
\end{subfigure}
\caption{\label{fig:schema_with_example} The JSON Schema is depicted as a graph, showing the hierarchy of properties for each data record. The example demonstrates the use of most properties including metadata for interpretability.}
\end{figure}

\newcommand{\ra}[1]{\renewcommand{\arraystretch}{#1}}
\begin{table*}
\centering
\ra{1.3}
\begin{tabular}{m{4.5cm}m{2.5cm}m{6cm}}\toprule
Field &  Type & Description \\
\midrule
\multicolumn{3}{l}{\textbf{Top-level attributes}}\\
\midrule
\texttt{network-type} (optional) & enum & "directed", "undirected", or "asc" \\
\texttt{metadata} (optional) & object & network-level metadata \\
\texttt{incidences} (required) & array & lists of records representing edge-node pairs \\
\texttt{nodes} (optional) & array & lists of node records \\
\texttt{edges} (optional) & array & lists of edge records\\[0.1in]
\multicolumn{3}{l}{\textbf{Entries in the \texttt{incidences} array}}\\
\midrule
\texttt{edge} (required) & string or integer & an edge ID \\
\texttt{node} (required) & string or integer & a node ID belonging to that edge \\
\texttt{weight} (optional) & number & incidence weight \\
\texttt{direction} (optional) & enum & "head" or "tail" \\
\texttt{attrs} (optional) & object & library-specific non-structural attributes\\[0.1in]
\multicolumn{3}{l}{\textbf{Entries in the \texttt{nodes} array}}\\
\midrule
\texttt{node} (required) & string or integer & a global node ID\\
\texttt{weight} (optional) & number & node weight \\
\texttt{attrs} (optional) & object & library-specific non-structural attributes\\[0.1in]
\multicolumn{3}{l}{\textbf{Entries in the \texttt{edges} array}}\\
\midrule
\texttt{edge} (required) & string or integer & a global hyperedge ID\\
\texttt{weight} (optional) & number & edge weight \\
\texttt{attrs} (optional) & object & library-specific non-structural attributes\\
\bottomrule
\end{tabular}
\caption{\label{tab:format} \textbf{A description of the HIF schema.} The schema is separated into five types of network data: the type of the higher-order network (encoded in \texttt{network-type}), network-level metadata (encoded in \texttt{metadata}), lists of nodes and edges with their attributes (encoded in \texttt{nodes} and \texttt{edges}, respectively), and structural information comprising information about the constituent nodes, hyperedges, and incidences.}
\end{table*}

The HIF specification organizes data about a higher-order network in five different categories via five top-level fields in the JSON file: \texttt{network-type}, \texttt{metadata}, \texttt{nodes}, \texttt{edges}, and \texttt{incidences} as can be seen in Table~\ref{tab:format}.
The text corresponding to the \texttt{network-type} field indicates the type of the higher-order network and offers additional guidance for processing the dataset which we will discuss in detail below.
The data corresponding to the \texttt{metadata} field is a dictionary-like object representing network-level attributes such as the name of the dataset, when it was collected, the author of the dataset, corresponding references, and other high-level information enabling greater dataset interpretability.
The \texttt{nodes} and \texttt{edges} fields store data records for the nodes and edges, respectively.
These records can be used to specify metadata corresponding to each node and hyperedge, or as we will discuss in more detail below, to specify empty edges or isolated nodes.
Lastly, the \texttt{incidences} field stores data records detailing node-hyperedge relationships (known as an \textit{incidence}).
Below we detail more of the structure as well as some of the implications of this specification.

We start with the \texttt{incidences} data, because this is the only top-level attribute which explicitly encodes the complex relationships between nodes via hyperedges.
The \texttt{incidences} field is the \textit{only} required field in HIF-compliant files, and a file only including this field corresponds to a higher-order network without any network, node, or hyperedge attributes.
The incidence data are composed of a list of entries known as \textit{records}, where each record indicates a relationship between a node and a hyperedge, known as an \textit{incidence}.
For example, \{"edge": 1, "node": 3\} indicates that node 3 belongs to hyperedge 1.
The collection of all records where "edge" is 1 forms a hyperedge, e.g., the collection of records
\{"edge": 1, "node": 3\}, \{"edge": 1, "node": 4\}, \{"edge": 1, "node": 7\}
corresponds to a hyperedge comprising the interaction between nodes 3, 4, and 7.
Additional structural features which can be specified are \texttt{weight} and, in the case of directed hypergraphs, \texttt{direction}.
These properties are added as top-level attributes in each record; e.g.,
\{"edge": 1, "node": 3, "weight": 2\}
specifies an incidence with an edge-dependent node weight of 2 and
\{"edge": 1, "node": 3, "direction": "tail"\}
indicates that node 3 is in the tail of directed hyperedge 1.
Incidences can also have properties, which indicate hyperedge-dependent nodal attributes.
All of these attributes are bundled as dictionary-like objects in each incidence record under the \texttt{attrs} field, e.g.,
\{"edge": 1, "node": 3, "attrs": \{"role": "PI"\}\}
might indicate node 3's role as a PI on project 1.

Similarly, we can specify attributes for nodes and hyperedges by creating records corresponding to the \texttt{nodes} and \texttt{edges} fields, respectively.
For example,
\{"node": 0, "attrs": \{"height": 176, "weight": 143\}\}
corresponds to node 0 with a height of 176 and weight of 143, and
\{"edge": 1, "attrs": \{"duration": 93, "setting": "coffee shop"\}\}
corresponds to an 93 minute interaction at a coffee shop.
Another use of these fields is to specify empty hyperedges and isolated nodes which don't participate in these incidence relationships and thus are not listed in the incidence records.
We can add these nodes and edges as node and edge records with or without attributes to ensure that they are present in the higher-order network.

The \texttt{network-type} field can take one of three values: "asc", "directed", and "undirected".
This field offers additional guidance for processing the higher-order network.
For example, using the "asc" keyword indicates that, even if all subfaces are not specified, that the network should be downward closed.
The best practice for storing an abstract simplicial complex is to only store the maximal faces and subfaces with attributes to minimize the storage required, but this is a guideline, not a hard requirement.
When using the "directed" keyword, this indicates that the library should expect the "direction" field indicating whether the incidence forms part of the head or tail of the directed hyperedge.
When the \texttt{network-type} is not specified, it is assumed that the type is an undirected hypergraph.

The \texttt{metadata} field can take arbitrary information to support the dataset under study. The schema allows any object in the \texttt{metadata} property, including deeply nested structures. Dataset authors are encouraged to use a flat object and to use only lowercase characters and hyphens, a convention commonly known as "dash case". This can be seen in Figure~\ref{fig:schema_with_example} where the "Creation Date" property name is encoded as "creation-date", and "DOI" is encoded as "doi".

\subsection{Checking HIF compliance}

A central contribution of the HIF project is a JSON schema allowing simple validation of datasets against the standard.
There are simple JSON validators in several different languages and we illustrate simple ways to validate datasets against the HIF Standard in Figure~\ref{fig:schema_validation_code}.

\begin{figure}
\centering
\vspace{0.5cm}
\textbf{Python}
\begin{lstlisting}[style=python]
import fastjsonschema
import json
import requests

schema = requests.get(url).json()
validator = fastjsonschema.compile(schema)
hiftext = json.load(open(filename,'r'))
try:
  validator(hiftext)
  print("HIF-Compliant JSON.")
except Exception as e:
   print(f"Invalid JSON: {e}")
\end{lstlisting}

\textbf{R}
\begin{lstlisting}[style=R]
library(jsonvalidate)
library(jsonlite)

schema <- paste(readLines(url, warn = FALSE))
validator <- json_validator(schema)
if (validator(filepath)) {
    print("HIF-Compliant JSON.")
} else {
    print("Invalid JSON.")
}
\end{lstlisting}

\textbf{Julia}
\begin{lstlisting}[style=julia]
using HTTP
using JSON3
using JSONSchema

schema = String(HTTP.get(url).body)
validator = Schema(schema)
hiftext = JSON3.read(filepath)
result = JSONSchema.validate(validator, hiftext)
println(
    result === nothing ?
    "HIF-Compliant JSON." :
    "Invalid JSON: " * "$result"
)
\end{lstlisting}
\caption{Example code snippets for validating the schema in different languages. In all cases, \texttt{url}=``https://raw.githubusercontent.com/pszufe/HIF-standard/main/schemas/hif\_schema.json'' and \texttt{filepath} is the local filepath of the file being validated.}
\label{fig:schema_validation_code}
\end{figure}

We emphasize that the validating the schema should be the first step in the implementation of an independent software packages which utilize this standard are ultimately responsible for reading and writing datasets to and from the appropriate network representation.
For example, the standard uses the "asc" keyword to specify that every interaction present in the hypergraph is downward closed, but checking the dataset against the schema doesn't explicitly check this.
In addition, best practices are that only the maximal simplex faces and subfaces with corresponding attributes are stored for abstract simplicial complexes, but this informal standard is not enforced by the schema.
Lastly, while end users are able to, in principle, specify the \texttt{direction} keyword for network datasets even if the \texttt{network-type} keyword is not \texttt{directed} and to omit the \texttt{direction} keyword even when \texttt{network-type} is \texttt{directed}, for simplicity, we simply say that \texttt{direction} field is optional.

\subsection{Updating the schema}

Changes made to the schema are documented in a \texttt{CHANGELOG.md} document.
When a new version is released, the version process is followed:
\begin{enumerate}
    \item Decide on the new version number based on the changes made since the last version and the Semantic Versioning guidelines~\autocite{preston-werner_semantic_2023}.
    \item Copy the schema in the \texttt{hif\us schema.json} file to a file named \texttt{hif\us schema\us <version>.json} where \texttt{version} indicates the new version.
    \item Add the changes made since the last release to \texttt{CHANGELOG.md} in a section with the new version as the name.
    \item Upload the new stable version to Zenodo as a persistent reference.
\end{enumerate}

The \texttt{hif\us schema.json} schema will always have version "latest" and all unit tests are based on this schema.

\section{Integration with software libraries}

A strength of the HIF standard is its integration with five prominent software libraries for higher-order network analysis: the Hypergraph Analysis Toolbox (HAT)~\autocite{pickard_hat_2023}, Hypergraphx~\autocite{lotito_hypergraphx_2023}, HyperNetX~\autocite{praggastis_hypernetx_2024}, SimpleHypergraphs.jl~\autocite{spagnuolo_analyzing_2020}, and XGI~\autocite{landry_xgi_2023}.
Here we detail the ways in which each library supports the HIF standard as well as brief descriptions of their use.

\subsection{Hypergraph Analysis Toolbox}

The \textbf{H}ypergraph \textbf{A}nalysis \textbf{T}oolbox (HAT)\footnote{https://hypergraph-analysis-toolbox.readthedocs.io}~\autocite{pickard_hat_2023} is a general-purpose software for constructing, visualizing, and analyzing hypergraphs and higher-order structures, with a focus on their structure and dynamics.
HAT implements tensor based methods for the analysis of higher-order observability/controllability~\autocite{pickard_observability_2023, pickard_geometric_2024}, coupling~\autocite{pickard_kronecker_2024}, and structural properties suitable for analysis of a wide range of data~\autocite{chen_tensor_2020, dotson_deciphering_2022}.
It provides HIF-compliant read and write capabilities in the HAT.Hypergraphs module.
Figure \ref{fig:hat-snippet} presents a sample using the HAT library.
The \texttt{Hypergraph.to\us hif} method converts the hypergraph object back to a Python dictionary according to the HIF schema, which can be saved to a JSON file.
\begin{figure}
\centering
\vspace{0.5cm}
\begin{lstlisting}[style=python]
import json
from HAT import Hypergraph
with open("example_hif.json") as file:
    hif_data = json.load(file)
HG = Hypergraph.from_hif(hif_data)
hif_output = HG.to_hif()
\end{lstlisting}
\caption{The example code demonstrates creating a HAT hypergraph from an example HIF JSON file and exporting it back to Python.}
\label{fig:hat-snippet}
\end{figure}

\subsection{Hypergraphx}

Hypergraphx (HGX)\footnote{https://github.com/HGX-Team/hypergraphx} is an open-source Python library designed for the analysis of complex systems characterized by higher-order interactions.
HGX offers an extensive collection of tools and algorithms for constructing, manipulating, analyzing, and visualizing hypergraphs, enabling users to capture the rich structure and study the dynamics of real-world systems beyond pairwise relationships.
HGX enables seamless storage and conversion of higher-order data across multiple representations, including hypergraphs, abstract simplicial complexes, bipartite graphs, line graphs and clique expansions, all while supporting hyperedges with weights, direction, sign, temporal dynamics and multiplexity.
The library provides tools for basic statistical characterizations (e.g., hyperedge size distribution), higher-order centrality, motif analysis (with scalable sampling), community detection, filtering, and synthetic hypergraph generation.
HGX also supports the simulation of dynamical processes and offers advanced visualization functionalities for exploring the structure of real-world systems.

On the data and I/O side, HGX is accompanied by \texttt{HGX-data}, a collection of ready-to-use datasets for higher-order network analysis in Python, including collaboration, face-to-face, biological data, and more.
HGX offers dedicated input and output functions, including a binary format to enable fast and efficient I/O over large-scale higher-order datasets.
Hypergraphx has integrated HIF support through the \texttt{read\us hif} function, allowing users to load hypergraphs directly from HIF-compliant JSON files and automatically handling the conversion to HGX’s internal data structures.
Similarly, the \texttt{write\us hif} function enables users to export an HGX \texttt{Hypergraph} object to the HIF format, preserving both the structural information and any associated node, edge, or incidence attributes.
This interoperability allows users to seamlessly leverage HGX functionalities while moving back and forth across different libraries.
In the future, HIF will accommodate additional hypergraph representations available in Hypergraphx, including multiplex hypergraphs~\autocite{lotito_multiplex_2024}. Moreover, HGX-data is planned to be fully HIF-compatible in order to be accessed by the whole higher-order software ecosystem.

An example demonstrating how to load a hypergraph from a HIF file, leverage HGX functionalities such as motif analysis, and export the data is provided in Figure~\ref{fig:hgx_example_hif}.
\begin{figure}
\centering
\vspace{0.5cm}
\begin{lstlisting}[style=python]
import hypergraphx as hgx

# Load HIF hypergraph data to HGX object
from hypergraphx.readwrite import read_hif
HG = read_hif(filename="example_hif.json")

# Compute and evaluate motifs 
from hypergraphx.motifs import compute_motifs
m = compute_motifs(HG, runs_config_model=20)

# Save HGX hypergraph object to HIF file
from hypergraphx.readwrite import write_hif
write_hif(HG, filename="example_hif.json")
\end{lstlisting}
\caption{Example demonstrating how to load a HIF-compliant file, use Hypergraphx (HGX) functionalities and export the file.}
\label{fig:hgx_example_hif}
\end{figure}

\subsection{HyperNetX}

HyperNetX (HNX) (\cite{praggastis_hypernetx_2024})\footnote{https://github.com/pnnl/HyperNetX} is a python package focused on developing tools for the analysis and visualization of hypergraphs.
HNX provides a user-friendly interface for creating, manipulating, and studying hypergraphs, with an emphasis on visualization capabilities that capture higher-order relationships. Some of the key features of HNX include:
\begin{itemize}
    \item Support for incidence, node, and hyperedge attributes, allowing for richer data representation.
    \item A suite of network analytical functions, including line graphs, degree sequences, connectivity measures, and centrality metrics specifically modified to be appropriate for hypernetworks (\cite{aksoy_hypernetwork_2020}).
    \item Support for topological measures of hypergraphs, including simplicial and zigzag homology of temporal hypergraphs (\cite{myers_topological_2023}).
    \item Tutorials for both basic and advanced hypergraph analysis methods with a contributors guide for the addition of new modules and tutorials.
    \item Visualization tools for rendering hypergraphs in various layouts, highlighting different aspects of their structure.
    \item Advanced analytical capabilities through supplementary modules including generative models, Laplacian clustering, and hypergraph modularity.

\end{itemize}

HyperNetX has integrated HIF through the \texttt{read\us hif} and \texttt{to\us hif} functions: \texttt{read\us hif} loads hypergraphs directly from HIF-compliant JSON files, handling the conversion to HNX's internal data structure and \texttt{to\us hif} exports an HNX hypergraph object to the HIF format, preserving both the structural information and any associated node, edge, or incidence attributes.
An example demonstrating how to load and save an HNX hypergraph is shown in Figure~\ref{fig:hnx_example_hif} where the filename is provided.
Alternatively, the JSON object can be loaded first and then the JSON object converted to an HNX hypergraph can be done.
\begin{figure}
\centering
\vspace{0.5cm}
\begin{lstlisting}[style=python]
import json
import hypernetx as hnx
# Load HIF hypergraph data to hnx object
HG = hnx.from_hif(filename = "example_hif.json")
# Save HIF data to file
hnx.to_hif(HG, filename="example_hif.json")
\end{lstlisting}
\caption{Example demonstrating how to load and export and HIF file using HyperNetX (HNX).}
\label{fig:hnx_example_hif}
\end{figure}

\subsection{SimpleHypergraphs.jl}

SimpleHypergraphs.jl\footnote{https://github.com/pszufe/SimpleHypergraphs.jl} is a Julia package designed to efficiently work with hypergraphs~\autocite{spagnuolo_analyzing_2020}.
Its key capabilities include (1) creating hypergraphs from incidence matrices, random hypergraph models, a Graphs.jl graph, or external data; (2) manipulating hypergraphs, (3) analyzing hypergraphs with fast algorithms for computing, for example, connected components or modularity based clustering; (4) visualizing hypergraphs; (5) integration with the Graphs.jl Julia package via bipartite graphs or a hypergraph projection; and (6) and file I/O.
SimpleHypergraphs.jl is designed to support (1) two-way of storage of node and hyperedge incidence information, which enables efficient traversal of the hypergraph in both directions --- from nodes to hyperedges and vice versa; (2) hyperedge-dependent node weights, i.e., incidence weights; (3) node and hyperedge metadata; (4) compatibility with Julia's \texttt{AbstractMatrix} interface, enabling efficient manipulation of hypergraphs using standard Julia matrix operations; and (5) lazy representation of hypergraph views which are compatible with the \texttt{AbstractGraph} interface of the Graphs.jl library, enabling seamless integration and direct usage of Graphs.jl algorithms.
Figure \ref{fig:simplehypergraphs_example_hif} presents an example of SimpleHypergraphs.jl library usage with HIF.
\begin{figure}
\centering
\vspace{0.5cm}
\begin{lstlisting}[style=julia]
using SimpleHypergraphs
hg = Hypergraph{Int, String, String}(
    [1 nothing       2 nothing; 
     3       1 nothing       4])
set_vertex_meta!(hg, "vertex 1", 1)
set_hyperedge_meta!(hg, "h-edge 2", 2)
g_save("hg_hif.json", hg; format=HIF_Format())
loaded_hg = hg_load(
    "hg_hif.json";
    format=HIF_Format(),
    HType=Hypergraph,
    T=Int, V=String, E=String
)
\end{lstlisting}
\caption{The example code demonstrates creating a hypergraph with node and hyperedge metadata represented as strings and exporting the hypergraph to a HIF-compliant JSON file.}
\label{fig:simplehypergraphs_example_hif}
\end{figure}
    
\subsection{XGI}

The Comple\textbf{X} \textbf{G}roup \textbf{I}nteractions (XGI)\footnote{https://xgi.readthedocs.io}~\autocite{landry_xgi_2023} software package provides support for hypergraphs, directed hypergraphs, and abstract simplicial complexes.
It provides a comprehensive suite of analysis and visualization tools, including algorithms, generative higher-order network models, linear algebra representations, conversions between many different data structures, and an integrated statistical interface.
Integrated with this software environment is the XGI-DATA repository, which hosts many higher-order datasets.
XGI provides not only visualization and analysis tools for higher-order networks but the ability to read and write numerous different file formats such as bipartite edge lists, hyperedge lists, and incidence matrices.

XGI currently supports the HIF standard in two ways: first, it provides HIF-compliant read and write capabilities and second, it provides HIF-compliant datasets in the XGI-DATA collection.
XGI has implemented an \texttt{hif} sub-module in its \texttt{readwrite} module, providing two methods: \texttt{read\us hif} and \texttt{write\us hif}.
These methods allow XGI to store undirected hypergraphs, directed hypergraphs, and abstract simplicial complexes according to the HIF standard.
When storing directed hypergraphs, XGI makes use of the \texttt{direction} field in the incidence records.
When storing abstract simplicial complexes, XGI only stores the maximal faces and hyperedges with associated attributes, with the expectation that the \texttt{asc} keyword will indicate that XGI should enforce downward closure when reading in the file.
An example demonstrating how to create a hypergraph, save the hypergraph as an HIF file, and then loading that file is shown in Figure~\ref{fig:xgi_example_hif}.
\begin{figure}
\centering
\vspace{0.5cm}
\begin{lstlisting}[style=python]
import xgi
H = xgi.Hypergraph(
    [[1, 2, 3], [2, 3, 4], [1, 4]]
)
example_file = "example_hif.json"
xgi.write_hif(H, example_file)
H = xgi.read_hif(example_file, nodetype=str)
H.edges.members()
[{"1", "2", "3"}, {"2", "3", "4"}, {"1", "4"}]
\end{lstlisting}
\caption{The code example demonstrates using XGI to read and write HIF-compliant files. First, a hypergraph with hyperedges $\{1, 2, 3\}$, $\{2, 3, 4\}$, and $\{1, 4\}$ is generated and written to an HIF-compliant JSON file. Second, the hypergraph is read from that file, casting the names of the nodes from integers to strings, and the hyperedge list is returned.}
\label{fig:xgi_example_hif}
\end{figure}

The \texttt{load\us xgi\us data} function in XGI supports reading HIF-compliant files from the XGI-DATA repository and several datasets in XGI-DATA are stored in XGI-DATA.
Future plans include converting all datasets to HIF, to make XGI-DATA a shared resource for all higher-order software libraries.

\section{Case study}

\begin{figure*}
    \centering
    \includegraphics[width=\linewidth]{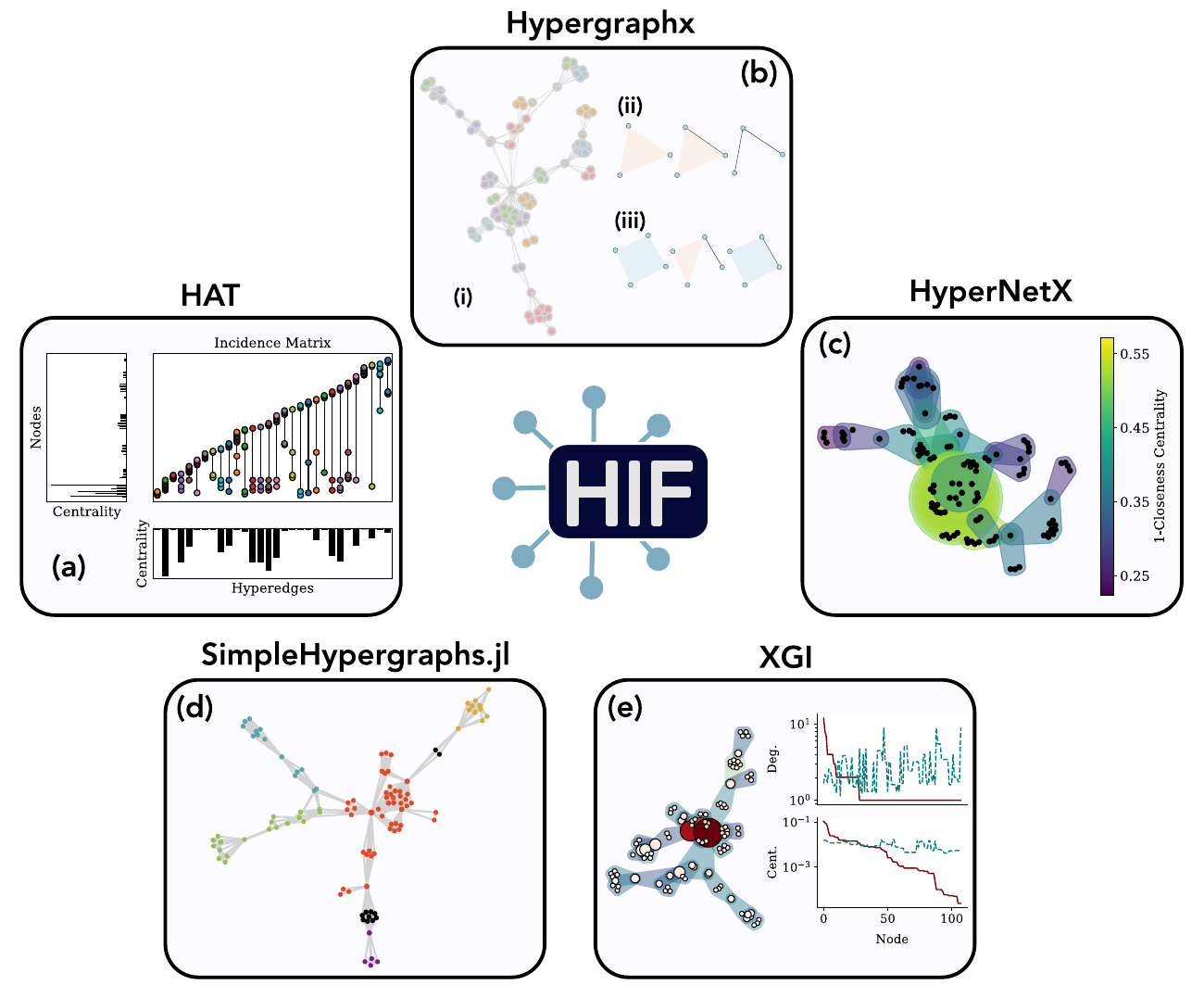}
    \caption{The Hypergraph Interchange Format enables interoperability between many popular higher-order network analysis libraries, as seen in this case study which analyzes a publication dataset. All code to reproduce this figure is available on the HIF-standard repository (See \href{https://github.com/pszufe/HIF-standard}{https://github.com/pszufe/HIF-standard}). The notebook that creates each visualization can be found in the \texttt{HIF-demo.ipynb} file. In the center, we display the HIF logo. In panel (a), we use HAT to compute the nonlinear eigenvector centralities of the nodes and hyperedges.
    In panel (b), we use Hypergraphx to visualize the largest connected component of the hypergraph (i), with node colors indicating the community structure. We also visualize the most frequent patterns of group interactions involving three (ii) and four nodes (iii). In panel (c), we use HyperNetX to solely examine the largest component of the hypergraph, compute the closeness centrality for each hyperedge, and color each hyperedge according to its centrality as we can see from the color bar. In panel (d), we use SimpleHypergraphs.jl to find the largest component, compute modularity-based community detection, and to visualize the network projected from the hypergraph, where each hyperedge becomes a clique. In panel (e), we use XGI to compute nodal statistics using the statistics interface; in the visualization, the node size is scaled by the nodal degree and the node color corresponds to the clique eigenvector centrality (CEC)~\autocite{benson_three_2019}. Similarly, in the top inset figure, we see the degree (solid line) and average neighbor degree (dashed line), while in the bottom inset figure, we see the CEC (solid line) and the H-eigenvector centrality (HEC)~\autocite{benson_three_2019,aksoy_scalable_2024} (dashed line). Note that the node labels are sorted in descending order according to the degree (top) and CEC (bottom).}
    \label{fig:HIF_case_study}
\end{figure*}

To illustrate the practical application and interoperability enabled by the Hypergraph Interchange Format (HIF), we present a case study.
Figure \ref{fig:HIF_case_study} provides an overview of how the HIF standard facilitates the exchange of hypergraph data across several hypergraph analysis software packages --- Hypergraphx, HyperNetX, SimpleHypergraphs.jl, and XGI --- enabling more integrated and sophisticated data analysis pipelines.

\subsection{Dataset}

We create the \texttt{publications.hif.json} HIF-compliant file and interact with this dataset by loading the file through each software library's API.

\begin{figure}[h!]
    \centering
    \includegraphics[width=0.9\linewidth]{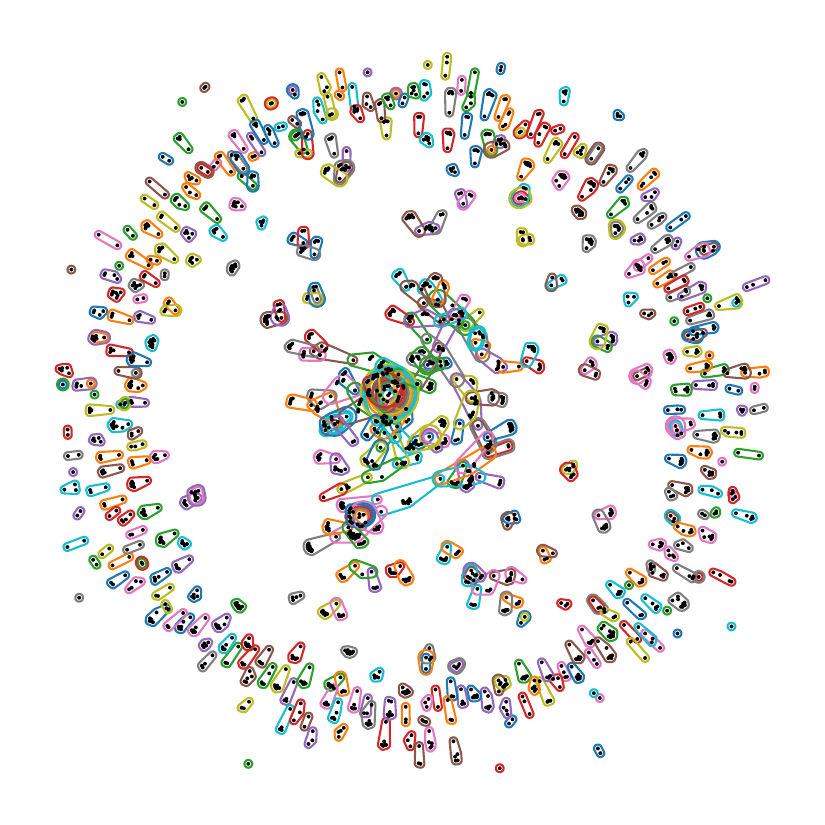} 
    \caption{Illustration of the full publications hypergraph including disconnected components.}
    \label{fig:hg_pubs_full}
\end{figure}
We use a publication dataset for our case study because of its rich attributes and tractable size.
This dataset consists of open-source publications with the keyword "Hypergraph" and was collected from ArXiv, BioRxiv, the DBLP computer science bibliography, and the U.S. Office of Scientific and Technical Information (OSTI).
The resulting hypergraph is composed of hyperedges, which represent scientific publications, and nodes, which represent authors and can be seen in Figure~\ref{fig:hg_pubs_full}.
Each hyperedge (publication) has a number of attributes including funding agencies, abstract, publication date, keyword tags, and the source, while each node (author) has an associated institution have the attributes of institutions (There are no incidence attributes).
The hypergraph has 1,960 nodes (authors) and 533 edges (publications).
A detailed description and the following analysis of this dataset using each of the hypergraph packages is provided in the notebook \texttt{HIF-demo.ipynb} in the tutorials folder of the HIF GitHub page \footnote{See \href{https://github.com/pszufe/HIF-standard}{https://github.com/pszufe/HIF-standard}.}.
We briefly describe examples of the types of analysis that one can perform on this dataset for each software package.

\subsubsection{Hypergraph Analysis Toolbox}

For HAT we demonstrate analyzing the publications hypergraph using the eigenvector centrality measure. This allows us to analyze node and edge centrality indicating the contributions of each author and publication. In Figure~\ref{fig:HIF_case_study}(a), the incidence matrix of the largest connected component of the hypergraph is shown, along with the node and hyperedge centrality measures. Each row of the incidence matrix indicates an author (node), and the columns of the matrix correspond to publications (hyperedges). HAT computes the centralities of the nodes and hyperedges, displaying them along the corresponding modes of the incidence matrix.

\subsubsection{Hypergraphx}

To demonstrate the capabilities of Hypergraphx, we consider our example publications dataset and perform an exploratory analysis combining community detection, motifs, and visualization.
Specifically, in Figure~\ref{fig:HIF_case_study}(b), we apply (overlapping) community detection to the largest connected component, identifying groups of closely related nodes~\autocite{contisciani_inference_2022}, and visualize the structure of the resulting hypergraph.
We further use Hypergraphx to mine the frequency of subhypergraph patterns of sizes 3 and 4~\autocite{lotito_exact_2024}, describing the hypergraph of publications at its microscale and providing insights into the structural building blocks of higher-order interactions present in the data~\autocite{lotito_higherorder_2022a}.
This use case illustrates how Hypergraphx can serve as a key component of an integrated workflow within the higher-order network ecosystem, enabling complex tasks such as motif analysis and community detection to be performed efficiently and in combination with other tools.

\subsubsection{HyperNetX}

For HyperNetX, we demonstrate the usefulness of analyzing the main component of the publications hypergraph using the $s$-closeness centrality measure for $s=1$, visualized as an Euler diagram in Figure~\ref{fig:HIF_case_study}(c).
We did not use the full hypergraph for this centrality analysis as it is too dense and disconnected to visually discern hyperedge centrality.
The $s$-closeness centrality measure shows the closeness of each hyperedge (in this case, a publication) based on the shared authors.
The color scale indicates the centrality value, with warmer colors (yellow) representing hyperedges that are more central, meaning they share more authors with other publications.
This analysis can highlight influential publications that are central to the collaborative structure of this publication hypergraph.

\subsubsection{SimpleHypergraphs.jl}

For the analysis with SimpleHypergraphs.jl we have selected the largest connected component of the citation hypergraph that has $108$ vertices and $30$ hyperedges.
Next, we use library's functionality hypergraph modularity-based clustering introduced in~\autocite{kaminski_clustering_2019} to identify clusters within the citation hypergraph.
Subsequently, we project the hypergraph to a pairwise representation where every hyperedge becomes a clique.
We visualize the graph in Figure~\ref{fig:HIF_case_study}(d), scaling node sizes by their degrees and using colors to denote community labels.
All communities with a single node are colored in black.

\subsubsection{XGI}

A strength of the XGI software package is its statistics interface, allowing practitioners to easily visualize, output as common Python data types, and compute statistics of node and edge properties.
In Figure~\ref{fig:HIF_case_study}(e), we see a hypergraph with its nodes colored and sized according to its degree and the average degree of its neighbors.
XGI can also easily filter datasets by hyperedge size according to the framework in~\autocite{landry_filtering_2024}.
Another strength of the XGI software package is its ability to quickly remove data artifacts with its \texttt{cleanup} method which has the ability to remove singletons, isolated nodes, multi-edges, and disconnected components.
In Figure~\ref{fig:HIF_case_study}(e), we see the largest component of the hypergraph without any multi-edges, isolated nodes, or singletons.

\section{Discussion}

In this paper, we introduced the Hypergraph Interchange Format as a standard for sharing hypergraph data across different scientific workflows.
We demonstrated its effectiveness with a case study showcasing the use of several software packages to analyze a single dataset in a variety of ways, each catering to a different library's strengths.
In the current fragmented open-source ecosystem, the HIF standard offers interoperability between many higher-order network analysis libraries.

As part of this effort, we surveyed the landscape of open-source software for existing standards and potential adopters, reaching out to more than 20 open-source projects with support for hypergraphs and abstract simplicial complexes.
Our exploration validated the hypothesis that the community of higher-order networks needs an open standard and that the HIF design supports many use cases.
Future work will include integration with more software libraries to improve its reach in the higher-order network science community.

The HIF standard currently emphasizes human readability and support for richly attributed data, but one can imagine more efficient ways for storing datasets which could be implemented in future iterations of HIF.
For example, Javascript Object Notation Lines (JSONL) is a specification defining how to transmit a top-level list of JSON objects as a series of one-line JSON documents, enabling streaming file input/output (I/O).
When solely considering performance, one could consider a database architecture with a query language such as GraphQL as a performant alternative for storing and retrieving higher-order network data.
As large higher-order datasets become increasingly more available, efficient storage and handling of higher-order network data will become more and more necessary.

Lastly, while the HIF standard directly supports undirected and directed hypergraphs and abstract simplicial complexes, support for additional higher-order representations will make HIF more widely useful.
Examples include hypergraphs with ordered and partially ordered sets (posets) which are useful in legal analysis~\autocite{coupette_legal_2024} and the Science of Science.
Several types of higher-order networks are implicitly supported by the HIF standard: signed hypergraphs which assign a sign +/- to each hyperedge; multilayer hypergraphs which assign nodes and hyperedges to different layers representing, for example, modes of transportation, relationship types, or social media platforms; and temporal hypergraphs where nodes and hyperedges have associated time intervals in which they are active.
The specific structural information necessary for each of these representations can be stored in the \texttt{attrs} field, but the schema excludes these as expected JSON fields.
In the future, adding explicit support for these fields will improve compatibility with libraries such as ASH~\autocite{failla_attributed_2023}, Raphtory~\autocite{steer_raphtory_2024}, and Reticula~\autocite{badie-modiri_reticula_2023}.
In addition, we recommend that when releasing new higher-order datasets that an HIF-compliant file is included in the release [for example, see~\autocite{smith_blue_2025a}].

While there are many avenues for future growth of HIF, we believe that this is a significant first step in the standardization of the higher-order network data sharing ecosystem.
It has a direct impact on the reproducibility of published works that adopt the standard.
HIF will facilitate seamless data exchange, cross-software collaboration, and more compact higher-order data analysis pipelines and will be a significant advance for the coming years of higher-order network science.

\paragraph{Funding Statement}
N.W.L. acknowledges support from the University of Virginia Prominence-to-Preeminence (P2PE) STEM Targeted Initiatives Fund, SIF176A Contagion Science.
This work acknowledges support from Pacific Northwest National Laboratory with information release number {PNNL-SA-217265}. Pacific Northwest National Laboratory is a multiprogram national laboratory operated for the US Department of Energy (DOE) by Battelle Memorial Institute under Contract No. DE-AC05-76RL01830. 

\paragraph{Competing Interests} None

\paragraph{Data Availability Statement}
All datasets, resources, documentation, and schemas are available on GitHub \footnotemark[\value{footnote}].
The stable schema is available on Zenodo~\autocite{zenodo_2025} and subsequent stable versions will be released on Zenodo.

\paragraph{Ethical Standards}
The research meets all ethical guidelines, including adherence to the legal requirements of the study country.

\paragraph{Contribution statement}

Author order is alphabetical.\\
Conceptualization: C.A.J., N.W.L., B.P., P.S.\\
Methodology: M.C., C.A.J., N.W.L., Q.F.L., A.M., J.P., B.P., P.S.\\
Project Administration and Supervision: N.W.L.\\
Software: M.C., N.W.L., Q.F.L., A.M., J.P., B.P., P.S.\\
Writing – Original Draft: M.C., C.A.J., N.W.L., Q.F.L., A.M., J.P., B.P., P.S.\\
Writing – Review \& Editing: M.C., C.A.J., N.W.L., Q.F.L., A.M., J.P., B.P., P.S.\\
Visualization: M.C., N.W.L., Q.F.L., A.M., J.P., P.S.
\printbibliography

\end{document}